\documentclass[12pt,a4paper]{article}

\usepackage{graphics}
\usepackage{latexsym}
\usepackage{amsmath}
\usepackage{amssymb}

\title{On CP violation and the measurement
of the dimuon charge asymmetry at hadron 
colliders}
\author{B. Hoeneisen}
\date{\small{
26 July 2006} }

\begin{document}
\maketitle

\begin{abstract}
\noindent
$B$ factories measure the CP violation parameter
of $B^0 \bar{B}^0$ mixing and decay.
Hadron colliders measure the dimuon charge asymmetry 
of an admixture of $B$ hadrons.
In this note we discuss a subtle point on 
how the CP violation parameter of 
$B^0_s \bar{B}^0_s$ mixing and decay can be 
extracted from these measurements.
\end{abstract}


\section{Introduction}
$B$ factories measure the CP violation parameter
of $B^0 \bar{B}^0$ mixing and decay.
Hadron colliders measure the dimuon charge asymmetry
of an admixture of $B$ hadrons.
In this note we discuss a subtle point on
how the CP violation parameter of
$B^0_s \bar{B}^0_s$ mixing and decay can be
extracted from these measurements.
This discussion is
particularly relevant since the parameters of mixing
and decay in the $B^0_s$ system, $\Delta M_s$,
$\Delta \Gamma_s$ and $\alpha_s \equiv \Re (\epsilon_{B^0_s})/ 
( 1 + \left| \epsilon_{B^0_s} \right|^2 )$
have come within experimental reach.

Let me summarize our arguments. We agree
with the derivation in \cite{PDG} of the decay
rates \textit{up to a normalization factor}.
This normalization factor determines the total
number of decays. For a \textit{given number} of decays,
the normalization factor may, or may not,
depend on CP violation. In this note we present in detail
two alternatives to illustrate this point.

The outline is as follows.
In Section 2 we briefly review the standard
formalism of $B^0 \bar{B^0}$ mixing and decay.\cite{PDG}
In Sections 3 and 4 we present two alternative
calculations of the time-integrated 
probabilities of $B$ and $\bar{B}$ to
decay to flavor specific final states. In Section 5
we present a discussion on these alternatives.
The difference between them is significant
for the $B^0_s$ system. 
Our conclusions are collected in Section 6.

\section{$B^0 \bar{B^0}$ mixing and decay}
Let us review the standard formalism of 
CP violation in mixing and decay.\cite{PDG} 
We take the Hamiltonian in
the ($\bar{B}^0, B^0)$,
or ($\bar{B}^0_s, B^0_s$), basis as
\begin{equation}
H \equiv M - \frac{i}{2} \Gamma \equiv
\left[ \begin{array}{cc}
m & M_{12} \\
M_{12}^* & m
\end{array}
\right] -
\frac{i}{2}
\left[ \begin{array}{cc}
\Gamma & \Gamma_{12} \\
\Gamma_{12}^* & \Gamma
\end{array}
\right],
\label{H}
\end{equation}
where the matrices $M$ and $\Gamma$ are hermitian.
The Hamiltonian $H$ itself is not hermitian since
the $B$ mesons do decay. This $2 \times 2$
Hamiltonian is an approximate description
of a system that has many more dimensions.

The solution of the Schr\"{o}dinger equation
$i \partial \psi / \partial t = H \psi$ with
\begin{equation}
\psi  \equiv \left( 
\begin{array}{c}
\bar{B}^0(t) \\
B^0(t)
\end{array} 
\right)
\label{psi}
\end{equation}
is
\begin{eqnarray}
\bar{B}^0(t) = \frac{1}{2} \left\{ s_+(t) + s_-(t) \right\} \bar{B}^0(0)
+ \frac{1 + \epsilon}{1 - \epsilon}
\cdot \frac{1}{2}
\left\{ s_+(t) - s_-(t) \right\}
B^0(0), \nonumber \\
B^0(t) = \frac{1 - \epsilon}{1 + \epsilon}
\cdot \frac{1}{2}
\left\{ s_+(t) - s_-(t) \right\} \bar{B}^0(0) +
\frac{1}{2} \left\{ s_+(t) + s_-(t) \right\} B^0(0),
\label{mixing}
\end{eqnarray}
where
\begin{eqnarray}
s_-(t)& = & \exp(-imt) \exp(-\Gamma t/2)
\exp(i \Delta M t/2) \exp(\Delta \Gamma t/4), \nonumber \\
s_+(t) & = & \exp(-imt) \exp(-\Gamma t/2)
\exp(-i \Delta M t/2) \exp(- \Delta \Gamma t/4),
\label{s}
\end{eqnarray}
\begin{equation}
\frac{1 + \epsilon}{1 - \epsilon} \equiv
\frac{\Delta M - \frac{i}{2} \Delta \Gamma}
{2 \left( M_{12}^* - \frac{i}{2} \Gamma_{12}^* \right)} =
\frac{2 \left( M_{12} - \frac{i}{2} \Gamma_{12} \right)}
{\Delta M - \frac{i}{2} \Delta \Gamma}.
\label{eps}
\end{equation}
The phase of $(1 + \epsilon)/(1 - \epsilon)$
is arbitrary: it can be changed by redefining the phase of
$\bar{B}^0(0)$. Observables depend on the absolute value
of $(1 + \epsilon)/(1 - \epsilon)$, or equivalently,
on the CP violation parameter of $B^0$ mixing and decay:
\begin{equation}
\alpha_q \equiv \frac{\Re(\epsilon)}
{1 + \left| \epsilon \right|^2},
\label{alpha}
\end{equation}
where $q = d, s$ (we write the subscript $q$ explicitly
only on quantities that are needed later).
For the same reason, we can multiply
$M_{12}$ and $\Gamma_{12}$ by a common phase-factor.
Only the relative phase is observable:
\begin{equation}
\angle \left\{ - \frac{\Gamma_{12}}{M_{12}} \right\} \equiv \phi.
\label{phase}
\end{equation}
We introduce the standard notation
\begin{equation}
x \equiv \frac{\Delta M}{\Gamma},
y \equiv \frac{\Delta \Gamma}{2 \Gamma}.
\label{xy}
\end{equation}

We now calculate the time integrated probabilities
for $B$ and $\bar{B}$ to decay to flavor specific
final states. Two alternatives are considered.

\section{Alternative A}
From (\ref{mixing}) we obtain the time integrated probability 
that a $\bar{B}^0_q$ decays as a $B^0_q$:
\begin{eqnarray}
\chi_q & = & \frac
{
\int_{0}^{\infty}{\left| \frac{1 + \epsilon}
{1 - \epsilon} \right|^2
\frac{1}{4} \left| s_+ - s_- \right|^2 dt}
}
{
\int_{0}^{\infty}{\left| \frac{1 + \epsilon}
{1 - \epsilon} \right|^2
\frac{1}{4} \left| s_+ - s_- \right|^2 dt }
+ \int_{0}^{\infty}{\frac{1}{4} \left| s_+ + s_- \right|^2 dt}
} .
\label{chibar1}
\end{eqnarray}
\textit{The denominator in (\ref{chibar1}) 
normalizes the probabilities}, so the
probability that a $\bar{B}^0_q$ decays as a  $\bar{B}^0_q$
is $1 - \chi_q$. This normalization of probabilities
\textit{defines} alternative A.

The integrals are
\begin{equation}
\int_{0}^{\infty} \left| s_+ \pm s_- \right|^2 dt
= \frac{2}{\Gamma} \left[ \frac{1}{1-y^2} \pm \frac{1}{1+x^2} \right],
\label{ss}
\end{equation}
so
\begin{eqnarray}
\chi_q & = & \frac{\left( x^2 + y^2 \right)
\left( \frac{1}{2} + \alpha_q \right)}
{1 + x^2 - 2 \alpha_q \left( 1 - y^2 \right)}.
\label{chi}
\end{eqnarray}
Similarly, the probability that a $B^0_q$ decays as a $\bar{B}^0_q$ is
\begin{equation}
\bar{\chi}_q = \frac{\left( x^2 + y^2 \right)
\left( \frac{1}{2} - \alpha_q \right)}
{1 + x^2 + 2 \alpha_q \left( 1 - y^2 \right) },
\label{chibar}
\end{equation}
and the probability that a $B^0_q$ decays as a
$B^0_q$ is $1 - \bar{\chi}_q$. These equations can be
found in \cite{bh}.

Let us now obtain the dimuon charge asymmetry 
in the limit $\alpha_q \ll 1$:
\begin{equation}
A_q \equiv \frac{N^{++} - N^{--}}{N^{++} + N^{--}}
= \frac{\chi_q (1 - \bar{\chi}_q) - \bar{\chi}_q (1 - \chi_q)}
{\chi_q (1 - \bar{\chi}_q) + \bar{\chi}_q (1 - \chi_q)} 
\approx 4 \alpha_q 
= \Im{\left\{ \frac{\Gamma_{12}}{M_{12}} \right\}}.
\label{dimu}
\end{equation}
In this limit, $\Delta M \approx 2 \vert M_{12} \vert$.
The single muon charge asymmetry is:
\begin{equation}
a_q \equiv \frac{N^+ - N^-}{N^+ + N^-} 
=\frac{\left[ \chi_q + (1-\bar{\chi}_q) \right] - \left[ \bar{\chi}_q + (1-\chi_q) \right]}
{\left[ \chi_q + (1-\bar{\chi}_q) \right] + \left[ \bar{\chi}_q + (1-\chi_q) \right]}
= \chi_q - \bar{\chi}_q
= A_q \xi_q,
\label{a}
\end{equation}
where $\xi_q \equiv \chi_q + \bar{\chi}_q - 2 \chi_q \bar{\chi}_q$.
The asymmetry of tagged \textquotedblleft{wrong-sign}"
decay rates induced by oscillations is equal to the dimuon charge asymmetry $A_q$:
\begin{eqnarray}
a_{SL}(t) \equiv \frac{\Gamma(\bar{B}^0_q \rightarrow B^0_q \rightarrow \mu^+\textrm{X})
- \Gamma(B^0_q \rightarrow \bar{B}^0_q \rightarrow \mu^-\textrm{X})}
{\Gamma(\bar{B}^0_q \rightarrow B^0_q \rightarrow \mu^+\textrm{X})
+ \Gamma(B^0_q \rightarrow \bar{B}^0_q \rightarrow \mu^-\textrm{X})}
\nonumber \\
= \frac{\left| \frac{1+\epsilon}{1-\epsilon} \right|^2 \left| s_+ - s_- \right|^2 
- \left| \frac{1-\epsilon}{1+\epsilon} \right|^2 \left| s_+ - s_- \right|^2}
{\left| \frac{1+\epsilon}{1-\epsilon} \right|^2 \left| s_+ - s_- \right|^2
+ \left| \frac{1-\epsilon}{1+\epsilon} \right|^2 \left| s_+ - s_- \right|^2}
\approx 4 \alpha_q.
\label{aSL}
\end{eqnarray}
Note that $a_{SL}(t)$ is independent of time $t$.

The preceding equations apply to the $B^0$,
or $B^0_s$, systems separately.
Let us now consider $N$ dimuon events with
a $B$ and a $\bar{B}$ hadron at production.
For simplicity
we consider only direct semileptonic decays $B \rightarrow \mu$X.
The sample has $N f_d$ mesons $B^0$,
$N f_s$ mesons $B^0_s$, and $N (1 - f_d - f_s)$ other
$B$ hadrons that do not mix.
Similarly, the sample has $N f_d$ mesons $\bar{B}^0$,
$N f_s$ mesons $\bar{B}^0_s$, and $N (1 - f_d - f_s)$ other    
$\bar{B}$ hadrons that do not mix.

The number of $\bar{B}$ hadrons that mix and decay as $\mu^+$X is
$N f_d \chi_d + N f_s \chi_s \equiv N \chi$.
So $\chi = f_d \chi_d + f_s \chi_s$. 
The number of $B$ hadrons that mix and decay as $\mu^-$X is
$N f_d \bar{\chi}_d + N f_s \bar{\chi}_s \equiv N \bar{\chi}$. 
So $\bar{\chi} = f_d \bar{\chi}_d + f_s \bar{\chi}_s$.
The number of $\bar{B}$ hadrons that decay as $\mu^-$X is
$N f_d (1 - \chi_d) + N f_s (1 - \chi_s) + N (1 - f_d - f_s) =
N (1 - \chi)$.
Finally, the number of $B$ hadrons that decay as $\mu^+$X is
$N f_d (1 - \bar{\chi}_d) + N f_s (1 - \bar{\chi}_s) + N (1 - f_d - f_s) =
N (1 - \bar{\chi})$.

The dimuon charge asymmetry is
\begin{equation}
A = \frac{\chi (1 - \bar{\chi}) - \bar{\chi} (1 - \chi)}
{\chi (1 - \bar{\chi}) + \bar{\chi} (1 - \chi)}
= f_d A_d \frac{\xi_d}{\xi} + f_s A_s \frac{\xi_s}{\xi},
\label{AA}
\end{equation}
where
$\xi \equiv \chi + \bar{\chi} - 2 \chi \bar{\chi}$.
The dimuon charge asymmetry can be written approximately as
\begin{equation}
A \approx f_d A_d \frac{\chi_{d0}}{\chi_0}
\cdot \frac{1 - \chi_{d0}}{1 - \chi_0} 
+ f_s A_s \frac{\chi_{s0}}{\chi_0}
\cdot \frac{1 - \chi_{s0}}{1 - \chi_0},
\label{AAA}
\end{equation}
where $\chi_0$ is the value of $\chi$ or $\bar{\chi}$ 
in the absence of CP violation, and similarly for $\chi_{q0}$.

\section{Alternative B}
Alternative B is the same as Alternative A, except that $\epsilon$
is set to zero in the denominator of (\ref{chibar1}).
In other words, the probabilities add up to 1 only in the
case of no CP violation, i.e. $\alpha_q = 0$. 
The time integrated probability that a $\bar{B}^0_q$ decays as a
$B^0_q$ is
\begin{equation}
\chi_q = \frac{1 + 2 \alpha_q}{1 - 2 \alpha_q} 
\cdot \frac{x^2 + y^2}{2 (1+x^2)}
\equiv \frac{1 + 2 \alpha_q}{1 - 2 \alpha_q} \cdot \chi_{q0}.
\label{}
\end{equation}
The probability that a $B^0_q$ decays as a
$\bar{B}^0_q$ is
\begin{equation}
\bar{\chi}_q = \frac{1 - 2 \alpha_q}{1 + 2 \alpha_q} \chi_{q0}.
\label{}
\end{equation}
The probability that a $\bar{B}^0_q$ decays as a
$\bar{B}^0_q$, and the probability that a $B^0_q$ decays as a
$B^0_q$, are equal to $(1 - \chi_{q0})$, \textit{independent of
CP violation.} 
This is the \textit{definition} of alternative B.

The dimuon charge asymmetry $A_q$ is:
\begin{equation}
A_q \equiv \frac{N^{++} - N^{--}}{N^{++} + N^{--}} 
= \frac{\chi_q (1-\chi_{q0}) - \bar{\chi}_q (1-\chi_{q0})}
{\chi_q (1-\chi_{q0}) + \bar{\chi}_q (1-\chi_{q0})} 
\approx 4 \alpha_q 
= \Im{\left\{ \frac{\Gamma_{12}}{M_{12}} \right\}},
\label{} 
\end{equation}
The single muon charge asymmetry is:
\begin{eqnarray}
a_q & = & \frac{N^{+} - N^{-}}{N^{+} + N^{-}}
=\frac{\left[ \chi_q + (1-\chi_{q0}) \right] - \left[ \bar{\chi}_q + (1-\chi_{q0}) \right]}
{\left[ \chi_q + (1-\chi_{q0}) \right] + \left[ \bar{\chi}_q + (1-\chi_{q0}) \right]}
\nonumber \\
& \approx & \frac{\chi_q - \bar{\chi}_q}{2} = A_q \chi_{q0}
\label{aq}
\end{eqnarray}
The asymmetry of tagged \textquotedblleft{wrong-sign}"
decay rates induced by oscillations is 
$a_{SL}(t) \approx 4 \alpha_q \approx A_q$, 
as for alternative A.
The dimuon charge asymmetry for the
admixture of $B$ hadrons is:
\begin{equation}
A = f_d A_d \frac{\chi_{d0}}{\chi_0}
+ f_s A_s \frac{\chi_{s0}}{\chi_0}.
\label{BBB}
\end{equation}
Equation (\ref{BBB}) can be found in \cite{Nir}.

\section{Discussion}

We agree with the derivation in \cite{PDG} of the decay
rates \textit{up to a normalization factor}.
This normalization factor determines the total
number of decays. For a \textit{given number} of decays,
the normalization factor may, or may not,
depend on CP violation.

Alternatives A and B differ \textit{only} in the
normalizing factor for the probabilities:
for alternative B we have set $\epsilon = 0$ 
in the denominator of (\ref{chibar1}).

Note that in both alternatives A and B
we obtain the same (well known) dimuon charge asymmetry
$A_q = 4 \alpha_q$, where $q = d, s$.
We also obtain the same (well known)
asymmetry of tagged \textquotedblleft{wrong-sign}"
decay rates induced by oscillations $a_{SL}(t) = 4 \alpha_q$.
However, the single muon charge asymmetries
$a_q$ are different, and the dimuon charge
asymmetries $A$, corresponding to an admixture
of $B$ hadrons, are different.

About 99\% of the branching fractions of 
$B^0$ and $B^0_s$ are to flavor specific final states.
Let us consider the approximation in which we neglect
the branching fraction to non flavor specific final
states (such as J/$\psi$Ks). In this approximation,
a $\bar{B}^0_q$ can decay either as a $B^0_q$ or as
a $\bar{B}^0_q$. Therefore the sum of the corresponding
probabilities is 1. If the probability of the former
is $\chi_q$, then the probability of the latter is
$(1 - \chi_q)$. This approximation is 
alternative A.

Now consider alternative B.
By construction, \textit{the normalizing factor is independent
of CP violation}. Therefore, the time integrated
probabilities for $\bar{B}^0_q \rightarrow \bar{B}^0_q$
and $B^0_q \rightarrow B^0_q$ \textit{are independent
of CP violation}. Consider a \textit{given number} $N$ of events with a 
$B^0_q \bar{B}^0_q$ pair at production.
The $N$ $\bar{B}^0_q$'s can decay either to flavor specific
final states as a $B^0_q$ or as
a $\bar{B}^0_q$, or to non flavor 
specific final states $f_{nfs}$.
Similarly, the $N$ $B^0_q$'s can decay either to flavor specific
final states as a $\bar{B}^0_q$ or as
a $B^0_q$, or to non flavor specific final states
$f_{nfs}$.
To be specific, consider an \textit{increase} of
the number of decays $\bar{B}^0_q \rightarrow B^0_q$
due to CP violation. Then, \textit{for a given} $N$,
there must be a corresponding
\textit{decrease} of the number of decays 
$\bar{B}^0_q \rightarrow f_{nfs}$.
Also, the number of $B^0_q \rightarrow \bar{B}^0_q$
\textit{decreases} by approximately the same amount
due to CP violation.
Then, there must be a corresponding
\textit{increase} of the number of decays
$B^0_q \rightarrow f_{nfs}$. 

In summary, if alternative B is the correct 
description of nature, CP violation in 
\textquotedblleft{wrong sign}" decays to flavor specific
final states must be \textit{exactly compensated} by 
CP violation in \textquotedblleft{mixing and decay}" 
to non flavor specific final states. 
Such compensation implies a delicate 
balance between many branching fractions to, 
and CP violation asymmetries of, flavor 
specific and non flavor specific final states.
To my knowledge, such relations have not been
established for the standard model and new physics.

\section{Conclusions}
$B$ factories measure $\alpha_d$
(equal to $\frac{1}{4} A_d$,
where $A_d$ is the dimuon charge asymmetry
of direct decays of $B^0 \bar{B}^0$'s,
independently of the alternative A or B).
Hadron colliders measure the dimuon charge
asymmetry $A$ of an
admixture of $B$ hadrons.
$A_s$ can be extracted from these measurements
using equation (\ref{AAA}) of alternative A,
or equation (\ref{BBB}) of alternative B.
These equations differ by factors
$(1 - \chi_{d0})/(1 - \chi_0) \approx 
(1 - 0.186)/(1 - 0.127) = 0.93$ for $A_d$,
and $(1 - \chi_{s0})/(1 - \chi_0) \approx 
(1 - 0.5)/(1 - 0.127) = 0.57$ for $A_s$.

Our conclusion is that \textit{we do not know}
how the normalizing factor depends on
CP violation. It depends on how well
CP violation to non flavor specific 
final states compensates CP violation to
flavor specific final states.
We have considered two bench mark
alternatives A and B. The difference
is relatively small for $A_d$, but
becomes important for the extraction of
$A_s$ from the measurement of
the dimuon charge asymmetry at hadron
colliders.

Alternative A is a well defined approximation.
Alternative B assumes a dubious cancelation
of CP violation in flavor specific and non
flavor specific decays. Until 
\textquotedblleft{theory catches up}", we
advocate using the dimuon charge asymmetry (\ref{AAA})
of alternative A, instead of (\ref{BBB}) of
alternative B, as it is a better defined
approximation, and results in a more 
conservative error of $A_s$.

\end{document}